\documentclass[11pt]{article}
\usepackage{amssymb}
\usepackage{latexsym}
\usepackage{amsmath}
\usepackage{graphicx}
\usepackage{listings}
\usepackage{cite}

\def\al{\alpha}
\def\L{\Lambda}
\def\l{\lambda}
\def\k{\kappa}
\def\g{\gamma}
\def\be{\beta}
\def\de{\delta}
\def\P{\Phi}
\title{\LARGE{Static dilatonic black hole with nonlinear Maxwell and Yang-Mills fields of power-law type}}
\author{M. M. Stetsko\footnote{E-mail: mstetsko@gmail.com}\
\\
  {\small Department for Theoretical Physics, Ivan Franko National University of Lviv,}\\
{\small 12 Drahomanov Str., Lviv, UA-79005, Ukraine
         }}
\begin{document}
\maketitle

{\abstract{Static spherically symmetric black hole solution is obtained in the framework of Einstein-dilaton theory with nonlinear Maxwell and Yang-Mills fields of power-law type. It is observed that black hole might have two horizons similarly as it takes place for linear gauge fields case. Thermodynamics of the black hole is studied, namely the behaviour of temperature is examined and the first law is written. The concept of extended phase space is also utilized, namely we have written and analyzed the equation of state for the black hole and examined the Gibbs free energy. The Gibbs free energy shows that there is the first order phase transition if the temperature or the pressure are below their critical values and coupling parameters are relatively small. Increasing of the coupling parameters might give rise to appearance of a domain with the zeroth order phase transition, the existence of the latter one is specific for other types of dilatonic black holes. Prigogine-Defay ratio has been calculated, it shows that at the critical point the phase transition is not exactly of the second order, it is closer to the glass-type phase transition since the Prigogine-Defay ratio is not equal to one.  }}

\section{Introduction}
Black Holes' Physics is a flourishing area of investigation for recent decades. Numerous black holes' solutions were derived and studied in the framework of standard General Relativity as well as within various its  modifications or alternative theories of Gravity. Very extensive attention is drawn (attracted) to the black holes with additional material fields, for instance gauge fields or some sorts of scalar fields. The scalar fields are important for studies in Cosmology where they might be utilized to explain for example the Dark Energy/Dark Matter issue or for some scenarios of Inflation phenomenon.

It is known that some types of scalar fields, namely the so-called dilaton and axion fields might be introduced in String Theory, in particular they might be introduced by the dimensional reduction procedure. The dilaton-type fields also come on the stage if the dimensional reduction procedure is utilized in the framework of higher dimensional General Relativity, namely if one tries to go from a higher dimensional theory to some effective lower dimensional one \cite{Goteraux_JHEP12}. The important fact we should point out here is that in this case the dilaton field would be minimally coupled to lower dimensional gravity, nonminimal coupling might take place if the higher dimensional theory differs from Einsteinian General Relativity, for instance if one takes into consideration Gauss-Bonnet term. If additional matter fields are included the dimensional reduction gives rise to appearance of a specific nonminimal coupling between dilaton and corresponding matter fields.

Dilaton black holes have been studied very actively for more than three decades \cite{Gibbons_NPB88}. Starting form early studies which emphasized on its String Theory origin \cite{Garfinkle_PRD91,Witten_PRD91,Kallosh_PRD92,Gregory_PRD93}, numerous works have been written where plenty of solutions were obtained and examined \cite{Rakhmanov_PRD94,Poletti_PRD94,Cai_PRD96,Gao_PRD04,
Yazadjiev_CQG05,Chan_NPB95,Astefanesei_PRD06,Mann_JHEP06,Kunz_PLB06,
Brihaye_CQG07,Charmousis_PRD09,Sheykhi_PRD07,Fernando_PRD09,
Sheykhi_PRD14,Hendi_PRD15,Dehyadegari_PRD17,Pedraza_CQG19,
Bravo_PRD18,Kord_PRD15}. Since as we have noted above in the presence of additional matter fields, in particular of the gauge fields, the dimensional reduction gives rise to some nontrivial coupling between the dilaton and these gauge fields, thus the gauge fields are usually taken into account. The most interesting case of the gauge fields is ordinary abelian field defined by Maxwell-type Lagrangian, although some nonlinear types of Lagrangians have been attracted some attention \cite{Sheykhi_PRD14,Hendi_PRD15,Dehghani_PRD19}. The interest to dilaton black holes is driven by a broad area of possible applications of corresponding solutions starting from some issues in cosmological scenario \cite{Zaslavskii_IJMPA03,Bergshoeff_CQG05,Kastor_CQG17} and up to their possible applications to holographic superconductivity \cite{Liu_JHEP10} or holographic QCD \cite{Li_JHEP11}.

In contrast with abelian field case black holes and in particular dilatonic black holes with nonabelian gauge fields were not so widely studied. The latter fact might be explained by more complicated structure of the Yang-Mills fields, especially if they are coupled to gravity and from the other side by narrower area of their possible application.  Black hole solution with nonabelian field was studied for the first time more than four decades ago \cite{Yasskin_PRD75} in the context of a dyonic black hole \cite{Kasuya_PRD82}. The interest to black holes with nonabelian fields renewed in the late eighties \cite{Bartnik_PRL88,Bizon_PRL90,Volkov_JETP89,Straumann_PLB90}, namely it was shown instability of corresponding solutions in asymptotically flat case, whereas the stable solutions were derived in AdS space-time. The existence of stable solutions stimulated the interest to black holes with nonabelian fields and since that time a lot of new solutions appeared and their various aspects were studied \cite{Torii_PRD95,Lavrelashvili_NPB93,Volkov_PRD96,Volkov_PRep98,
Mavromatos_JMP98,Winstanley_CQG99,Bjoraker_PRL00,Kleihaus_PRD96,
Brihaye_PRD02,Brihaye_PLB03,Bij_PLB02,Brihaye_PRD07,Radu_CQG05,
Manvelyan_PLB09,Kleihaus_PRL03,Lerida_PRD09,Ghosh_PLB11,Mann_PRD06,Cvetic_PRD10,
Mazhari_PRD07,Mazhari_PLB08,Mazhari_GRG10,Mazhari_PRD11,Mazhari_EPJC13,
Bostani_MPLA10,Balakin_PRD16,Kleihaus_CQG16,Hendi_PLB18,Mazhari_PLB09,Ali_PRD19,Baxter_PRD07,Mazhari_PRD08}.
It should be pointed out that to derive the black holes' solutions various ansatze were applied, which allowed to obtain analytical or numerical solutions. Among the simplest one is the so-called magnetic Wu-Yang ansatz \cite{Mazhari_PRD07,Mazhari_PRD08,Mazhari_PLB08,Mazhari_GRG10,Bostani_MPLA10} which gives rise to magnetically charged solutions of relatively simple form. Dilatonic black holes with nonabelian fields were considered in \cite{Lavrelashvili_NPB93,Kleihaus_PRD96,Radu_CQG05,Mazhari_GRG10,
Mann_PRD06,Kleihaus_PRL03,Kleihaus_CQG16,Stetsko_EYM20,
Stetsko_EMYMD20,Stetsko_EnYM20}. 

In this work we investigate Einstein-dilaton theory coupled to nonlinear gauge fields, namely Maxwell and Yang-Mills fields with nonlinearities of power-law type and obtain a black hole's solution, as far as we know such kind of a theory has not been studied in the literature yet. It is known that linear Maxwell Lagrangian is conformally invariant in four dimensional case, but for other dimensions of space-time it loses its conformal invariance. Keeping of the conformal invariance for the Maxwell field Lagrangian was the main motivation to introduce the power-law nonlinearity of the Lagrangian \cite{Hassaine_PRD07} and as it was shown the power-law type of the field Lagrangian is the only form which maintains the conformal invariance if the power of nonlinearity equals to one fourth of the dimension of space-time. Immediately after the power-law conformally invariant gauge field Lagrangian was considered due to its relative simplicity power-law type Lagrangians with arbitrary power of nonlinearity were examined \cite{Hassaine_PRD07,Maeda_PRD09,Gonzalez_PRD09,Olmo_PRD11,Jing_JHEP11}. It should be also pointed out that the power or the parameter of nonlinearity is not completely arbitrary, to become physically acceptable the power-law Lagrangian should give rise to some physically reasonable behaviour of metric and gauge field, satisfy energy conditions and causality condition \cite{Maeda_PRD09,Gurtug_PRD19} and as a result if some particular nonlinear Lagrangian fulfills all the mentioned requirements the corresponding theory becomes consistent.  Similar conclusion about conformal invariance is valid also for the Yang-Mills field \cite{Mazhari_PLB09}. It should be also pointed out that power-law dependences for gauge field Lagrangians although with some other powers of nonlinearity appear in the framework of Euler-Heisenberg electrodynamics which is of quantum field theory origin \cite{Yajima_PRD01,Ruffini_PRD13} and in low energy limit of heterotic string theory \cite{Gross_NPB87}.    The model we consider here is a natural generalization of linear fields model examined in our earlier work \cite{Stetsko_EMYMD20}, from the other side in \cite{Stetsko_EnYM20} we studied Einstein-power-Yang-Mills-dilaton theory, thus the present work might also be treated as a generalization of the former model. Similarly as in our previous works \cite{Stetsko_EYM20,Stetsko_EMYMD20,Stetsko_EnYM20} the gauge group for the Yang-Mills filed is chosen to be $SO(n)$, and for the nonabelian gauge potential we use Wu-Yang ansatz, the latter allows us to derive relatively simple analytical solution.

The structure of this paper is as follows. In the second section we consider the action for our model, derive corresponding equations of motion and obtain static black hole's solution. In the third section we study some aspects of thermodynamics in the standard framework, namely we obtain, analyze the black hole's temperature and write the first law. In the fourth section the thermodynamics of the black hole is examined within the extended phase space concept, where the cosmological constant $\L$ is supposed to be a thermodynamic value, related to pressure. In this section we obtain and investigate the equation of state for the black hole and the Gibbs free energy, we also calculate Prigogine-Defay ratio which allows to characterize phase transition at the critical point. Finally, the fifth section contains some conclusions.

\section{Einstein-power-Maxwell-power-Yang-Mills-dilaton theory and black hole's solution}
The action for Einstein-power-Yang-Mills-power-Maxwell-dilaton theory takes the form:
\begin{eqnarray}\label{action}
 S=\frac{1}{16\pi}\int_{{\cal M}} {\rm d}^{n+1}x\sqrt{-g}\left(R-\frac{4}{n-1}\nabla^{\mu}\Phi\nabla_{\mu}\Phi-V(\Phi)-e^{-\frac{4\de\Phi}{n-1}}\left(Tr(F^{(a)}_{\mu\nu}F^{(a)\mu\nu})\right)^p+e^{-\frac{4\be\Phi}{n-1}}\left(-\cal{F}_{\mu\nu}\cal{F}^{\mu\nu}\right)^s\right)+S_{B},
\end{eqnarray}
where $g$ is the determinant of the metric tensor, $R$ is the Ricci scalar, $\P$ and $V(\P)$ denote dilaton field and dilaton potential respectively, $F^{(a)}_{\mu\nu}$ and $\cal{F}_{\mu\nu}$ are the Yang-Mills and Maxwell fields correspondingly, $\de$ and $\be$ are the dilaton-Yang-Mills and dilaton-Maxwell coupling constants respectively, $p$ and $s$ are nonlinearity parameters for Yang-Mills and Maxwell fields correspondingly. The second term $S_{B}=\frac{1}{8\pi}\int_{\partial{\cal M}} d^{n}x\sqrt{-h}K$ is the boundary Gibbons-Hawking-York term, where $K$ is the extrinsic curvature and $h$ is determinant of the boundary metric.

The Yang-Mills field tensor is defined as follows:
\begin{equation}\label{YM_field}
F^{(a)}_{\mu\nu}=\partial_{\mu}A^{(a)}_{\nu}-\partial_{\nu}A^{(a)}_{\mu}+\frac{1}{2\bar{\sigma}}C^{(a)}_{(b)(c)}A^{(b)}_{\mu}A^{(c)}_{\nu},
\end{equation} 
where $A^{(a)}_{\mu}$ is the Yang-Mills potential, $C^{(a)}_{(b)(c)}$ denotes the structure constants for the corresponding gauge group which, similarly as in our previous works is chosen to be $SO(n)$ and $\bar{\sigma}$ is the Yang-Mills coupling constant.

The Maxwell field tensor is defined in a usual manner:
\begin{equation}\label{Max_field}
{\cal{F}}_{\mu\nu}=\partial_{\mu}{\cal{A}}_{\nu}-\partial_{\nu}{\cal{A}}_{\mu},
\end{equation}
where ${\cal A}_{\mu}$ is the Maxwell potential component.

Now we write the equations of motion for the system described by the action (\ref{action}):
\begin{eqnarray}\label{einstein}
\nonumber R_{\mu\nu}=\frac{g_{\mu\nu}}{n-1}\left(V(\Phi)+(1-2p)e^{-\frac{4\de\Phi}{n-1}}\left(Tr(F^{(a)}_{\rho\sigma}F^{(a)\rho\sigma})\right)^p+(2s-1)e^{-\frac{4\be\Phi}{n-1}}\left(-\cal{F}_{\rho\sigma}\cal{F}^{\rho\sigma}\right)^s\right)\\
+\frac{4}{n-1}\partial_{\mu}\Phi\partial_{\nu}\Phi+2pe^{-\frac{4\de\Phi}{n-1}}
Tr(F^{(a)}_{\rho\sigma}{F^{(a)\rho\sigma}})^{p-1}Tr(F^{(b)}_{\mu\l}{F^{(b)\l}_{\nu}})+2se^{-\frac{4\be\Phi}{n-1}}\left(-\cal{F}_{\rho\sigma}\cal{F}^{\rho\sigma}\right)^{s-1}\cal{F}_{\mu\l}{\cal{F}_{\nu}}^{\l};
\end{eqnarray}
\begin{equation}\label{scal_eq}
\nabla_{\mu}\nabla^{\mu}\Phi=\frac{n-1}{8}\frac{\partial V}{\partial \Phi}-\frac{\de}{2}e^{-\frac{4\de\Phi}{n-1}}\left(Tr(F^{(a)}_{\rho\sigma}F^{(a)\rho\sigma})\right)^{p}-\frac{\be}{2}e^{-\frac{4\be\P}{n-1}}\left(-{\cal F}_{\rho\sigma}{\cal F}^{\rho\sigma}\right)^s;
\end{equation}
\begin{equation}\label{YM_eq}
\nabla_{\mu}\left(e^{-\frac{4\de\Phi}{n-1}}\left(Tr(F^{(d)}_{\rho\sigma}F^{(d)\rho\sigma})\right)^{p-1}F^{(a)\mu\nu}\right)+\frac{1}{\bar{\sigma}}e^{-\frac{4\de\Phi}{n-1}}\left(Tr(F^{(d)}_{\rho\sigma}F^{(d)\rho\sigma})\right)^{p-1}C^{(a)}_{(b)(c)}A^{(b)}_{\mu}F^{(c)\mu\nu}=0.
\end{equation}
\begin{equation}\label{em_eq}
\nabla_{\mu}\left(e^{-\frac{4\be\Phi}{n-1}}\left({\cal F}_{\rho\sigma}{\cal F}^{\rho\sigma}\right)^{s-1}{\cal{F}}^{\mu\nu}\right)=0.
\end{equation}
In the present work we consider a static black hole's solution, thus the metric can be cast in the form:
\begin{equation}\label{metric}
 ds^2=-W(r)dt^2+\frac{dr^2}{W(r)}+r^2R^2(r)d\Omega^2_{n-1},
\end{equation}
and here $d\Omega^2_{n-1}$ is the line element of $n-1$--dimensional unit hypersphere.

To define the gauge potential for the Yang-Mills field we utilize Wu-Yang ansatz, which allows us to write the gauge potential as follows:
\begin{equation}\label{gauge_pot}
{\bf A}^{(a)}=\frac{\bar{q}}{r^2}C^{(a)}_{(i)(j)}x^{i}dx^{j}, \quad r^2=\sum^{n}_{j=1}x^2_j,
\end{equation}
where indices $a$, $i$ and $j$ run the following ranges $2\leqslant j+1<i\leqslant n$ and $1\leqslant a\leqslant n(n-1)/2$, the new parameter $\bar{q}=\bar{\sigma}$ is chosen to be equal to the Yang-Mills coupling constant $\bar{\sigma}$. The coordinates $x_i$ in the upper relation might be defined in the form:
\begin{eqnarray}
\nonumber x_1=r\cos{\chi_{n-1}}\sin{\chi_{n-2}}\ldots\sin{\chi_1},\quad x_2=r\sin{\chi_{n-1}}\sin{\chi_{n-2}}\ldots\sin{\chi_1},\\
\nonumber  x_3=r\cos{\chi_{n-2}}\sin{\chi_{n-3}}\ldots\sin{\chi_1},\quad x_4=r\sin{\chi_{n-2}}\sin{\chi_{n-3}}\ldots\cos{\chi_1},\\
\nonumber \cdots \quad\quad\\
x_n=r\cos{\chi_1},
\end{eqnarray}
and the angular variables vary in the ranges $0\leqslant\chi_{i}\leqslant\pi, i=1,\ldots,n-2$, $0\leqslant\chi_{n-1}\leqslant 2\pi$. 
The angular variables $\chi_1,\ldots,\chi_{n-1}$ also allow to define the line element of the unit hypersphere in the metric (\ref{metric}), namely we write:
\begin{equation}
d\Omega^2_{n-1}=d\chi^2_{1}+\sum^{n-1}_{j=2}\prod^{j-1}_{i=1}\sin^2{\chi_{i}}d\chi^2_{j}.
\end{equation}

Using the potential (\ref{gauge_pot}) and calculating the gauge field $F^{(a)}_{\mu\nu}$ defined by the relation (\ref{YM_field}) one can verify that the equations (\ref{YM_eq}) are satisfied. We point out here that since we have utilized the Wu-Yang ansatz for the Yang-Mills field, the invariant for this field takes the same form as for the linear field case \cite{Stetsko_EYM20,Stetsko_EMYMD20}:
\begin{equation}
Tr(F^{(a)}_{\rho\sigma}F^{(a)\rho\sigma})=(n-1)(n-2)\frac{q^2}{r^4R^4}.
\end{equation}

\begin{figure}
\centerline{\includegraphics[scale=0.33,clip]{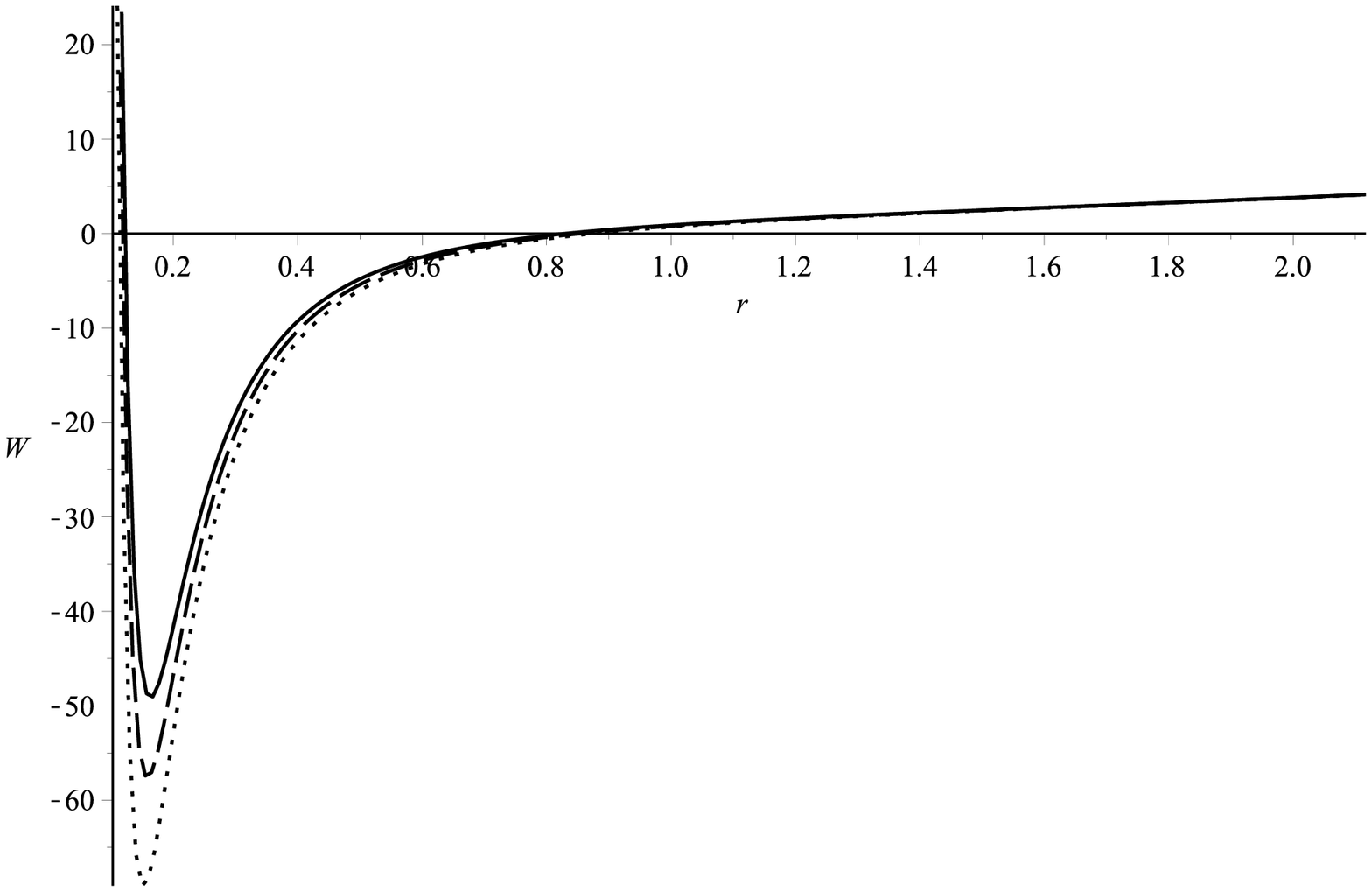}\includegraphics[scale=0.33,clip]{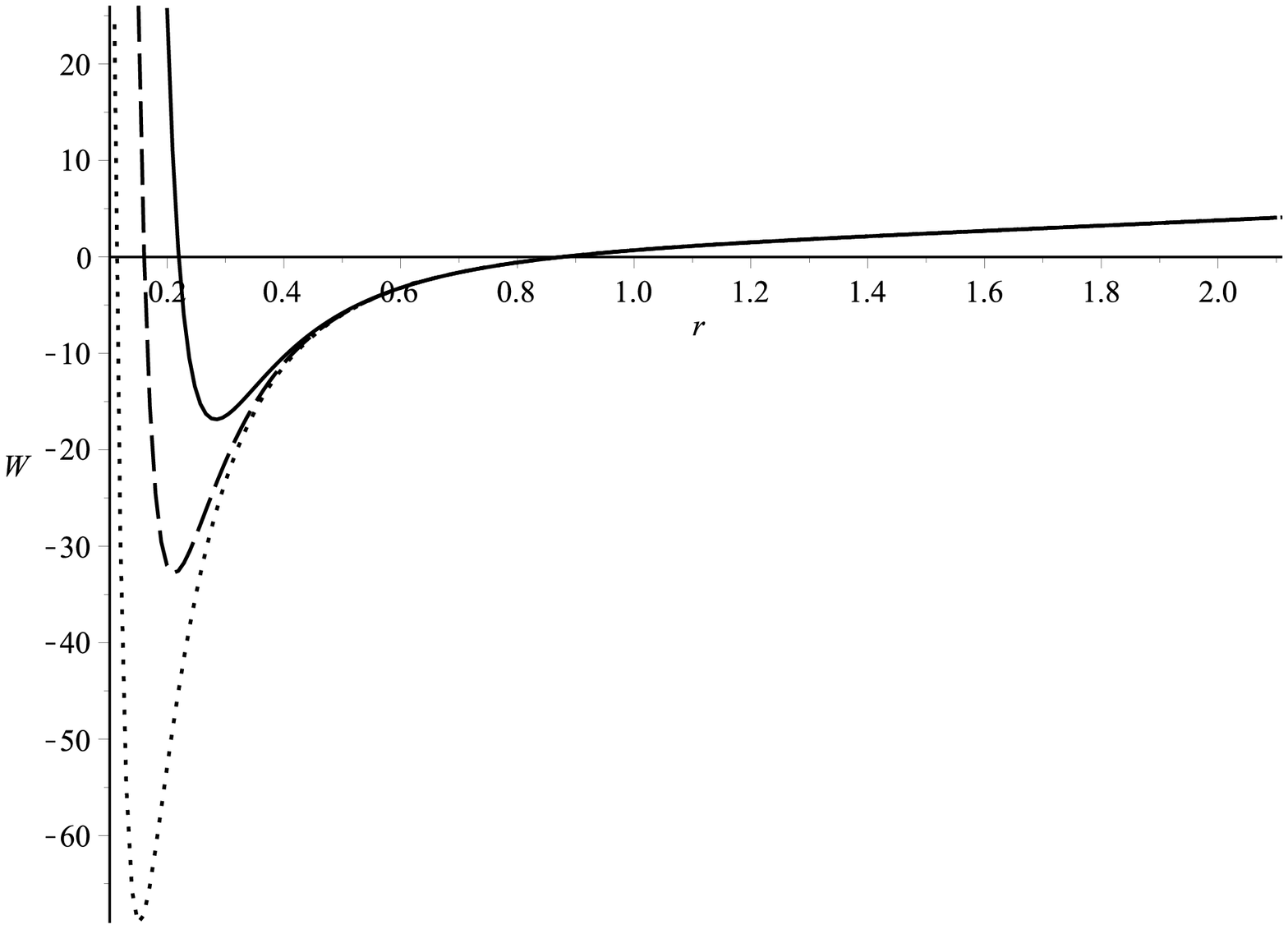}}
\caption{Metric function W(r) for various values  of nonlinearity parameters $p$ (the left graph) and $s$ (the right one). For both of the graphs we have taken: $n=5, m=1, b=1, \L=-12, \al=0.3, \de=0.2, \be =0.15, \bar{q}=0.2, q=0.3$. Variating parameters for the left graph are chosen to be $p=1$ (solid line), $p=1.15$ (dashed line), $p=1.2$ (dotted line) and $s=1.2$. For the right graph we have taken $s=1$, $s=1.1$ and $s=1.2$ which correspond to the solid, dashed and dotted curve respectively and here $p=1.2$.}\label{W_graph}
\end{figure}

We examine a static black hole's solution, thus the potential for the Maxwell field can be chosen in the form ${\cal{A}}={\cal{A}}(r)dt$. Having used the equation (\ref{em_eq}) and taking into account the evident form of the metric we obtain the electromagnetic field tensor:
\begin{equation}
{\cal F}_{rt}=\frac{q}{(rR(r))^{\frac{n-1}{2s-1}}}e^{\frac{4\be\P}{(n-1)(2s-1)}}.
\end{equation}
The invariant for the electromagnetic field takes the form:
\begin{equation}
{\cal F}_{\rho\sigma}{\cal F}^{\rho\sigma}=-2\frac{q^2}{(rR(r))^{\frac{2(n-1)}{2s-1}}}e^{\frac{8\be\P}{(n-1)(2s-1)}}.
\end{equation}
To solve equations of motion (\ref{einstein})-(\ref{scal_eq}) one should define the dilaton potential $V(\P)$. Here we use the so-called Liouville form for this potential:
\begin{equation}\label{dil_pot}
V(\Phi)=\Lambda e^{\lambda\Phi}+\Lambda_1 e^{\lambda_1\Phi}+\Lambda_2e^{\lambda_2\Phi}+\Lambda_3e^{\lambda_3\Phi},
\end{equation} 
and parameters $\l$, $\l_i$ and $\L_i$ might be derived when one tries to satisfy the equations of motion (\ref{einstein})-(\ref{scal_eq}). To solve the equations of motion we also make the following ansatz for the function $R(r)$:
\begin{equation}\label{ansatz_R}
R(r)=e^{2\al\Phi/(n-1)},
\end{equation} 
where $\al$ is some constant parameter. We point out here that similar ansatz  was utilized in numerous papers where black holes with dilaton fields were examined \cite{Chan_NPB95, Cai_PRD96,Sheykhi_PRD07,Stetsko_EYM20,Stetsko_EPJC19}, but in more recent of them \cite{Sheykhi_PRD07,Stetsko_EYM20,Stetsko_EPJC19} the parameter $\al$ in the function (\ref{ansatz_R}) was chosen to be equal to the dilaton coupling constant. We also note that in our previous work \cite{Stetsko_EMYMD20}, where we had two coupling constants, the parameter in the ansatz for the function $R(r)$ was chosen to be equal to one of them. Thus, the parameter $\al$ might be considered on equal footing with the dilaton field coupling parameters. Taking into account the written above ansatz (\ref{ansatz_R}) we might solve the equations of motion (\ref{einstein})-(\ref{scal_eq}) and we write:
\begin{gather}
\nonumber W(r)=-mr^{1+(1-n)(1-\gamma)}+\frac{(n-2)(1+\al^2)^2}{(1-\al^2)(\al^2+n-2)}b^{-2\gamma}r^{2\gamma}-\\\nonumber \frac{\Lambda(1+\al^2)^2}{(n-1)(n-\al^2)}b^{2\gamma}r^{2(1-\gamma)}+\frac{2p(n-1)^{p-1}(n-2)^{p}(1+\al^2)^2\bar{q}^{2p}}{(\al^2-2p+\al\de)(n+\al^2-4p+2\al\de)}b^{-2(2p+\k_1)\g}r^{2(1+\k_1\g)+4p(\g-1)}\\+\frac{2^{s}s(2s-1)^2(1+\al^2)^2q^{2s}}{(n+2\al\be-2s(1+\al^2)+\al^2)(s(n-1)+\al\be-(2s-1)\al^2)}b^{\frac{2(\k_2-s(n-1))\g}{2s-1}}r^{2\left(1+\frac{s(n-1)(\g-1)-\k_2\g}{2s-1}\right)},\label{W_nmym}
\end{gather}
where $\g=\al^2/(1+\al^2)$, $\k_1=\de/\al$, $\k_2=\be/\al$ and
\begin{equation}\label{phi_f}
\Phi(r)=\frac{\al(n-1)}{2(1+\al^2)}\ln{\left(\frac{b}{r}\right)}.
\end{equation}
It is worth pointing out that the parameter $m$ in (\ref{W_nmym}) is an integration constant related black hole's mass and the parameter $b$ in the upper two relations is the other integration constant, related to some rescaling properties. The parameters of the dilaton potential $V(\P)$ take the form as follows:
\begin{gather}
\l=\frac{4\al}{n-1},\quad \l_1=\frac{4}{\al(n-1)}, \quad \l_2=\frac{4(2p-\al\de)}{\al(n-1)},\quad \l_3=\frac{4(s(n-1)+\al\be)}{\al(n-1)(2s-1)};\\
\L_1=\frac{(n-1)(n-2)\al^2}{\al^2-1}b^{-2},\quad \L_2=-\frac{\al(\al+\de)}{\al^2-2p+\al\de}\left((n-1)(n-2)\right)^p\bar{q}^{2p}b^{-4p},\\\L_3=-\frac{\al(2s-1)(\be-\al(2s-1))}{\al\be+s(n-1)-(2s-1)\al^2}2^sq^{2s}b^{\frac{2s(1-n)}{2s-1}}.
\end{gather}
The parameter $\L$ in the dilaton potential (\ref{dil_pot}) might take arbitrary value and it is treated as the cosmological constant. We point out here that for linear fields case, namely when $p=s=1$ and if $\de=\al$ the metric function $W(r)$ (\ref{W_nmym}) and the parameters of dilaton potential are reduced to the corresponding values derived in our previous work \cite{Stetsko_EMYMD20}. Similarly if we do not take the Maxwell field into account (or if we put $q=0$) and $\de=\al$ our result is reduced to the result from our previous work \cite{Stetsko_EnYM20} and in case of linear Yang-Mills theory, the results of the work \cite{Stetsko_EYM20} are recovered. On the other hand if we eliminate the Yang-Mills field, namely if we impose $\bar{q}=0$,  the metric function $W(r)$ (\ref{W_nmym}) will be similar to the spherically symmetric case of the metric derived in the paper \cite{Kord_PRD15}, even though to have complete coincidence with the results of that work the coupling constant $\beta$ should be redefined. If the Yang-Mills field is excluded and in addition we impose $s=1$ (linear Maxwell field) and set $\alpha=\beta$ the obtained relation for the metric function completely coincides with corresponding spherically symmetric solution derived in \cite{Sheykhi_PRD07}. Although the obtained solution is a generalization of the results obtained in our earlier work \cite{Stetsko_EMYMD20} the main qualitative features of the obtained solution (\ref{W_nmym}) are very similar to linear fields case (if $p=s=1$), namely for large distances the leading term of the metric function is of the order $\sim\L r^{2(1-\g)}$ making the space-time asymptotically nonflat. It is worth emphasizing that asymptotic behavior for large $r$ ($r\to\infty$) is defined by the parameter $\al$ which in general is not equal to any of the dilaton-gauge fields coupling parameters $\be$ and $\de$. For small distances the leading terms are caused  by one of the gauge fields terms, namely it depends on the dimension $n$ and nonlinearity parameters $p$ and $s$. We point out here that for linear fields system $p=s=1$, the leading term is caused by the Maxwell field, but if $n=3$ and the parameter $\al$ is small, the Yang-Mills term might be comparable to the Maxwell field one. The behaviour  of the metric function $W(r)$ is shown on the Figure [\ref{W_graph}].  This figure reflects the main features of the metric function $W(r)$ mentioned above. As it is easy to see from the Figure [\ref{W_graph}] the metric function $W(r)$ might have two horizons, defined as roots of the equation $W(r_h)=0$, and the larger from them, which we denote as $r_+$ is the event horizon of the black hole. It should be pointed out here that if the charge parameters for the gauge field which is dominant for small distances goes up the horizons of the black hole become closer and they merge for some value of this parameter, in this case we have extreme black hole. In case this parameter is increased further instead of the black hole we have a naked singularity. We would like to stress that this kind of behaviour of the solution is not a some peculiarity related to the dilaton field or any kind of nonlinearity, similar situation takes place even for ordinary Reissner-Nordstrom solution, where the increase of the charge firstly gives rise to the extreme solution with the following appearance of a naked singularity. In addition it might be shown that the only point of physical singularity is the origin, whereas the horizon points where the metric (\ref{metric}) also shows singular behaviour are points of a coordinate sinularity as it usually takes place for black holes' solutions. 

\section{Thermodynamics of the black hole}
To derive the first law of black hole's thermodynamics we calculate the temperature of the black hole. It is known that the temperature can be defined with help of surface gravity and the final expression for the temperature takes the form:
\begin{gather}
\nonumber T=\frac{W'(r_+)}{4\pi}=\frac{(1+\al^2)}{4\pi}\left[\frac{(n-2)}{(1-\al^2)}b^{-2\g}r^{2\g-1}_{+}\left(1+\frac{2p(1-\al^2)}{\al^2-2p+\al\de}\times\right.\right.\\\left.\left.\nonumber\left((n-1)(n-2)\right)^{p-1}\bar{q}^{2p}b^{-2(2p-1+\k_1)\g}r^{2\left((1-2p)(1-\g)+\k_1\g\right)}_{+}\right)-\frac{\L}{n-1}b^{2\g}r^{1-2\g}_{+}-\right.\\\left.\frac{2^{s}s(2s-1)}{s(n-1)+\al\be-(2s-1)\al^2}q^{2s}b^{\frac{2(\k_2-s(n-1))\g}{2s-1}}r^{1+\frac{2(s(n-1)(\g-1)-\k_2\g)}{2s-1}}_{+}\right].\label{Temp}
\end{gather}
It is worth emphasizing that the relation (\ref{Temp}) is reduced to corresponding relations obtained in our previous works, namely if $p=s=1$ and $\de=\al$ we have the relation obtained for linear fields case \cite{Stetsko_EMYMD20}. If the Maxwell field is not taken into account, or we just set $q=0$ and put $\de=\al$ we arrive at the relation for power-law Yang-Mills case \cite{Stetsko_EnYM20} and when in addition we set $p=1$ the relation for linear dilaton-Yang-Mills black hole is recovered \cite{Stetsko_EYM20}. On the other hand if we eliminate the Yang-Mills field putting $\bar{q}=0$ the relation (\ref{Temp}) will be in agreement with corresponding relations for spherically symmetric solutions derived for nonlinear ($s\neq 1$) \cite{Kord_PRD15} and linear ($s=1$) \cite{Sheykhi_PRD07} Maxwell field respectively and their complete coincidence takes place after additional redefinition of coupling constants. Here we also would like to stress that since the metric function $W(r)$ (\ref{W_nmym}) and the temperature (\ref{Temp}) are reduced to corresponding relations in the mentioned above particular cases of linear fields or the absence any of them, similar consequences we will have for other thermodynamic relations which we will derive below, for instance for the equation of state, critical values, Gibbs free energy. 

The relation (\ref{Temp}) has rather complicated form, but some important features follow directly from its evident expression. Namely, since we take into account the cosmological constant $\L$ it means that for large radius of the horizon the corresponding term in the relation (\ref{Temp}) defines the behaviour of the temperature and in the limit when $\al\to 0$ the linear dependence, typical for AdS case is recovered. If $r_+$ is small the temperature decreases and its character is mainly defined by the corresponding gauge field term. For some intermediate values of the radius of horizon the temperature shows nonmonotonous behaviour which gives a hint about some kind of critical behaviour and a few aspects of which will be considered below.

\begin{figure}
\centerline{\includegraphics[scale=0.33,clip]{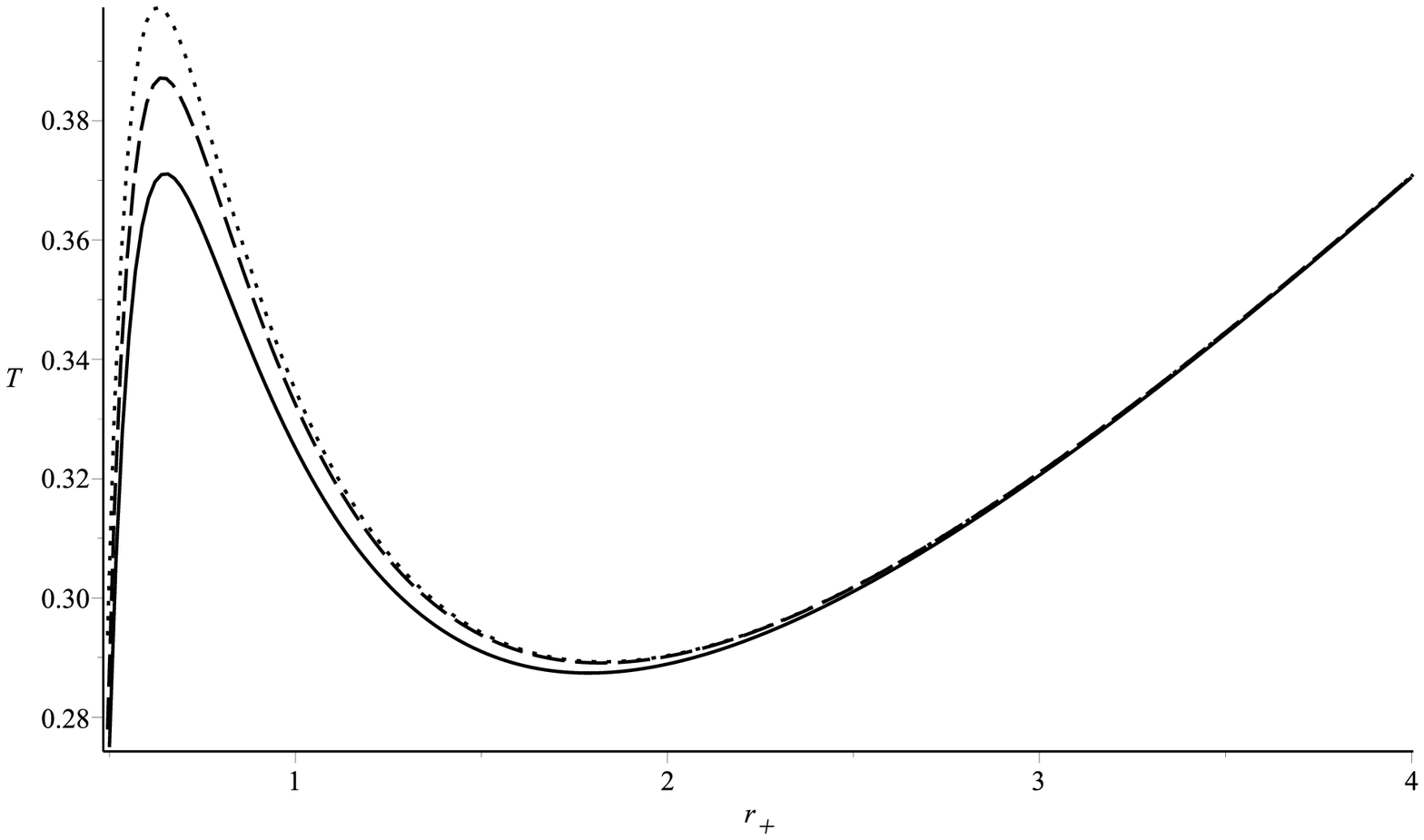}\includegraphics[scale=0.33,clip]{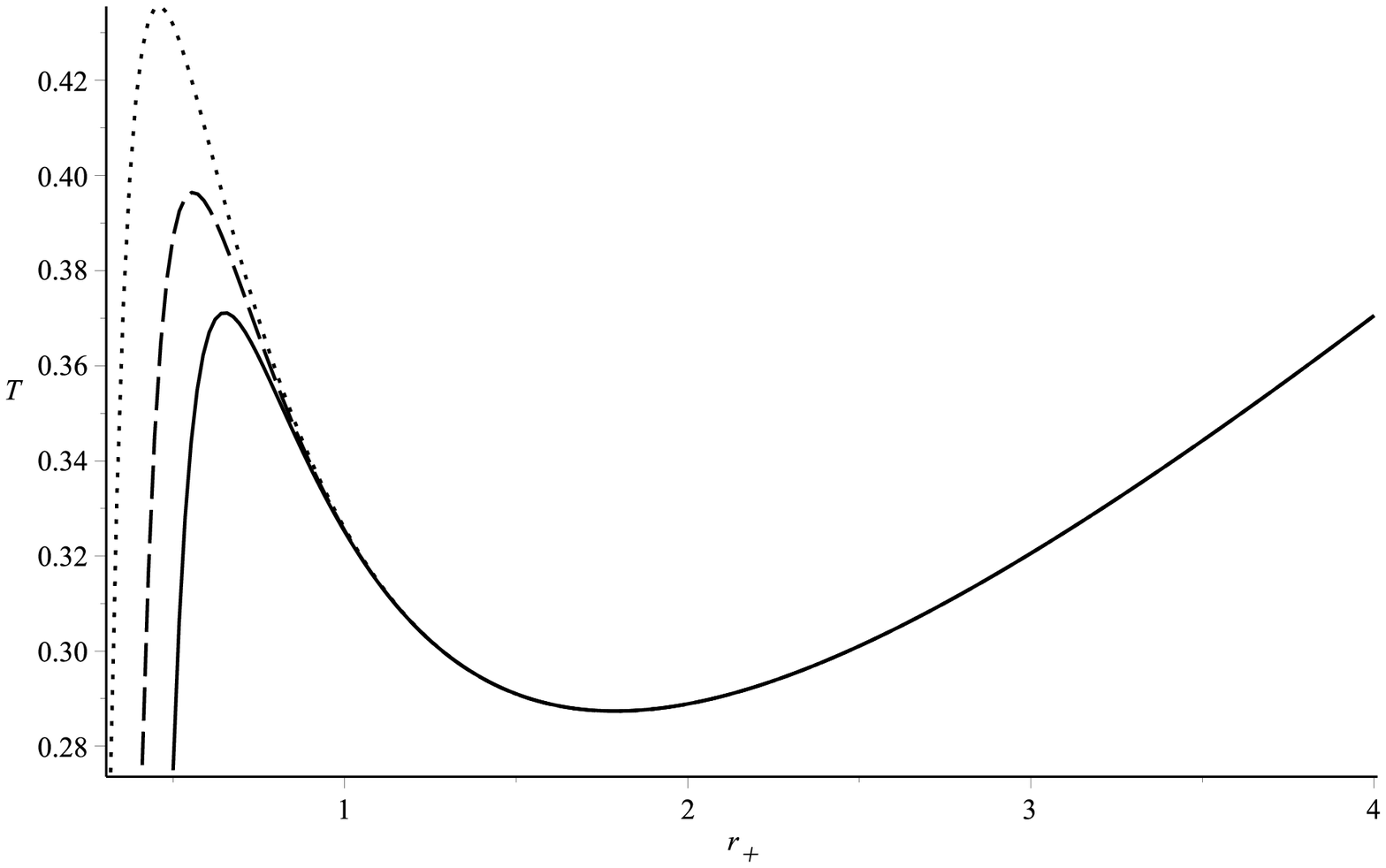}}
\caption{Black hole's temperature $T$ as a function of horizon radius  $r_+$ for various values of the parameters $p$ (the left graph) and $s$ (the right one). The fixed parameters for both graphs are: $n=5,\al=0.18, \de=0.12, \be=0.15, \L=-4, b=1, \bar{q}=q=0.2$. For the left graph the correspondence is as follows: the solid, dashed and dotted curves correspond to $p=1, p=1.4$ and $p=2$ respectively and here $s=1$. For the right graph we have: the solid, dashed and dotted curves correspond to $s=1, s=1.2$ and $s=1.5$ respectively and $p=1$.}\label{temp_graph}
\end{figure}

\begin{figure}
\centerline{\includegraphics[scale=0.33,clip]{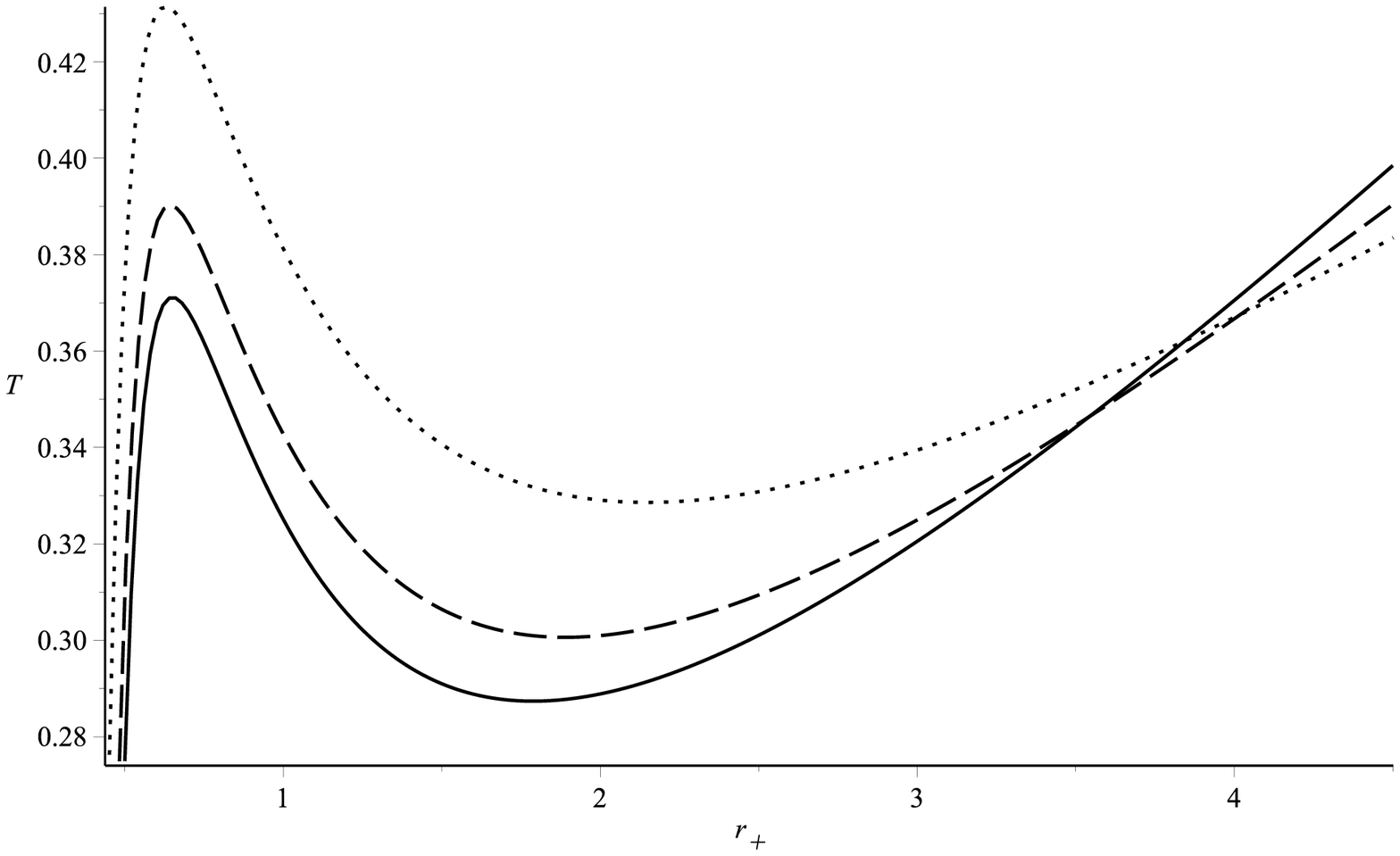}\includegraphics[scale=0.33,clip]{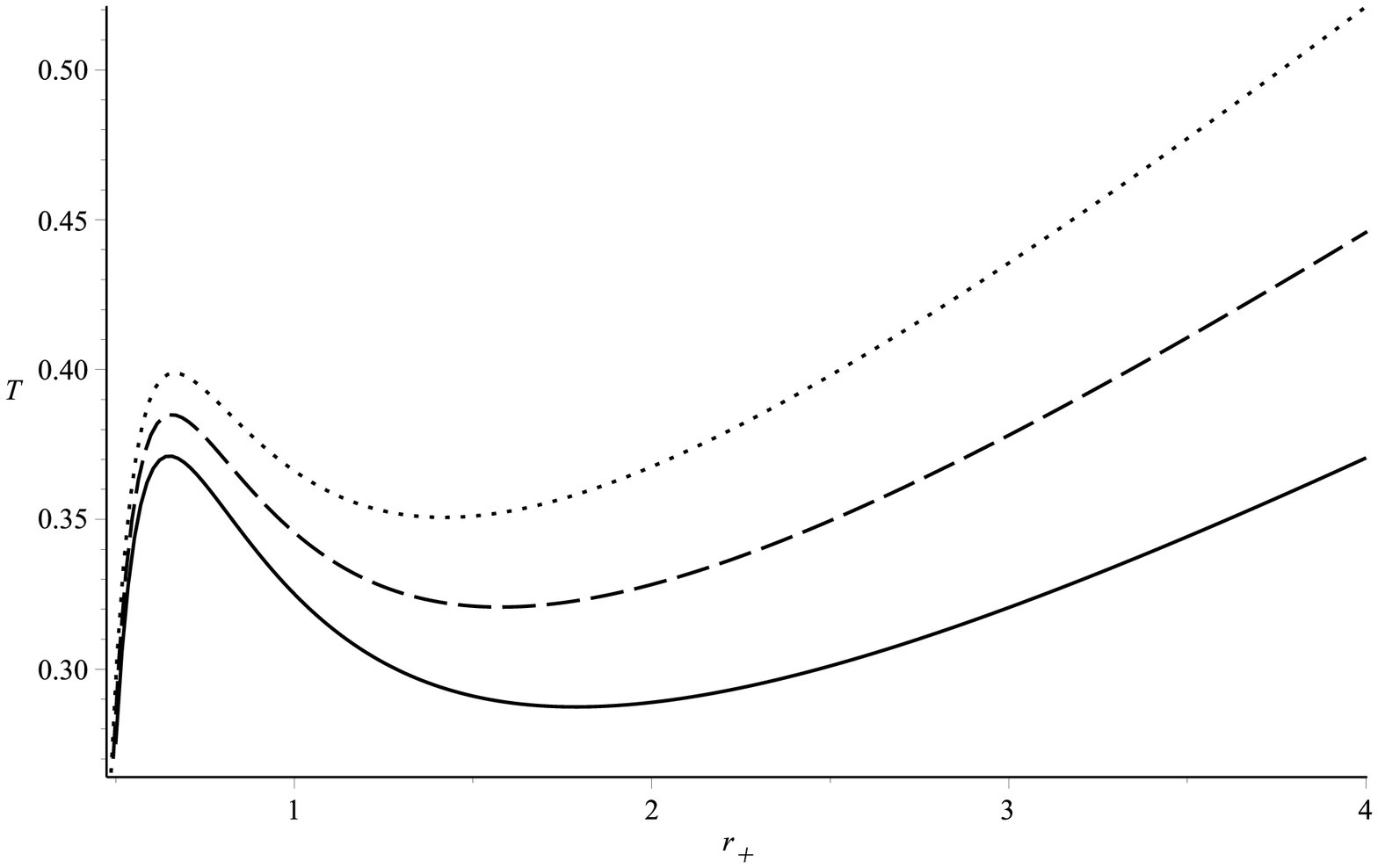}}
\caption{Black hole's temperature $T$ as a function of horizon radius  $r_+$ for various values of the parameter $\al$ (the left graph) and the cosmological constant $\L$ (the right one). The fixed parameters for both graphs are: $n=5, \de=0.12, \be=0.15, b=1, \bar{q}=q=0.2, s=1$. On the left the graph the solid, dashed and dotted curves correspond to $\al=0.18, \al=0.25$ and $\al=0.35$ respectively while $\L=-4$. On the right graph the solid, dashed and dotted lines correspond to $\L=-4$, $\L=-5$ and $\L=-6$ respectively and $\al=0.18$.}\label{temp_gr_2}
\end{figure}

 To comprehend the behaviour of the temperature (\ref{Temp}) better we show it graphically, what is demonstrated on the Figures [\ref{temp_graph}] and [\ref{temp_gr_2}]. Namely, the Figure [\ref{temp_graph}] shows the influence of nonlinearities on the temperature. We can conclude that increase of any of the nonlinearity parameters ($p$ or $s$) makes the peak on the dependence $T=T(r_+)$ higher. If the parameter $s$ rises this peaks also has a shift to the left, whereas the increase of the parameter $p$ does not show the shift so clearly. The Figure [\ref{temp_gr_2}] demonstrates the influence of the variation of the parameter $\al$ and the cosmological constant $\L$ on the temperature. We see that with increase of the parameter $\al$ makes the temperature higher for smaller and intermediate values of $r_+$, while for large radius of the horizon the temperature goes up slowly than for smaller values of $\al$. The manner the cosmological constant affects on the temperature is simpler, since the latter one depends linearly on the former. So the temperature increases on all the range of variation of the horizon radius $r_+$. It is worth noting here that for intermediate values of $r_+$ the function $T=T(r_+)$ becomes less monotonous  with increase of the absolute value of $\L$. Here we also point out that the variation of parameters $\k_1$ and $\k_2$ (or the coupling parameters $\de$ and $\be$ respectively) affects on the temperature in the range of small radii of the horizon since the coupling constants are given in the terms related to the gauge fields and those terms are substantial for small horizon radius. 

Since the gravitational part of the action (\ref{action}) is defined by the standard Ricci scalar and coupling between gravity and other fields is minimal the entropy of the black hole is defined as a quarter of the horizon area:
\begin{equation}\label{entropy}
S=\frac{\omega_{n-1}}{4}b^{(n-1)\g}r^{(n-1)(1-\gamma)}_+,
\end{equation}
and here $\omega_{n-1}$ is the surface area of a $n-1$--dimensional unit hypersphere. We point out here that the realtion (\ref{entropy}) might be derived in several ways, in particular by virtue of the Wald's entropy relation, which is applicable in our case \cite{Stetsko_EMYMD20}. 

Here we suppose that the electric charge of the black hole might be be varied and it means that it affects on thermodynamics of the black hole and we have to consider corresponding values, namely the electric charge as a thermodynamic one. The electric charge can be derived due to the relation:
\begin{equation}\label{charge_int}
Q=\frac{1}{4\pi}\int e^{-\frac{4\be\Phi}{n-1}}\left(-{\cal F}_{\rho\sigma}{\cal F}^{\rho\sigma}\right)^{s-1}*{\cal F}
\end{equation} 
where $*{\cal F}$ is the dual to the electromagnetic field form. It should be noted that the integral in the upper relation is taken over a space-like hypersurface enclosing the black hole. Having calculated the latter integral we arrive at:
\begin{equation}\label{charge}
Q=\frac{\omega_{n-1}}{4\pi}2^{s-1}q^{2s-1}.
\end{equation}
Since for our black hole solution (\ref{W_nmym}) other parameters are held fixed we can write the first law of black hole's thermodynamics in the following form:
\begin{equation}\label{1st_law}
dM=TdS+\Phi_e dQ,
\end{equation}
where:
\begin{equation}
T=\left(\frac{\partial M}{\partial S}\right)_Q, \quad \Phi_e=\left(\frac{\partial M}{\partial Q}\right)_S,
\end{equation}
and in the latter relation $\Phi_e$ is the electric potential which can be measured by an observer at the infinity. Having used the first law (\ref{1st_law}) one can derive the thermodynamic mass of the black hole which takes the form:
\begin{equation}\label{mass_bh}
M=\frac{(n-1)b^{(n-1)\gamma}\omega_{n-1}}{16\pi(1+\al^2)}m.
\end{equation} 
In the framework of ordinary thermodynamics one can also examine thermal stability of the black hole, for instance it can be done if one investigates the heat capacity $C_{Q}=T\left(\partial S/\partial T\right)_{Q}$. We point out here that  similar task was performed in our earlier work \cite{Stetsko_EMYMD20}, where linear fields were considered. From the qualitative point of view, at least when nonlinearity parameters $p$ and $s$ are close to one, the behaviour of the heat capacity would be similar to the linear fields case. This fact can be explained by similar behaviour of the temperature $T=T(r_+)$ and the same form of the expression for the entropy (\ref{entropy}). 

\section{Thermodynamics of the black hole in the extended phase space}
To obtain the extended phase space it is assumed that the cosmological constant $\L$ might be treated as a thermodynamic variable \cite{Kastor_CQG09,Cvetic_PRD11,Dolan_CQG11}, namely the negative cosmological constant plays the role of pressure. Since we take into consideration the dilaton field the thermodynamic pressure can be defined as follows \cite{Stetsko_EYM20,Stetsko_EMYMD20}:
\begin{equation}\label{press}
P=-\frac{\Lambda}{16\pi}\left(\frac{b}{r_+}\right)^{2\gamma}.
\end{equation}
In the extended framework the mass of the black hole $M$ is identified with the enthalpy function $M=H$. The thermodynamic volume which is conjugate to the pressure might be written in the forrm:
\begin{equation}\label{TD_volume}
V=\left(\frac{\partial H}{\partial P}\right)_S=\left(\frac{\partial M}{\partial P}\right)_S=\frac{\omega_{n-1}(1+\alpha^2)}{n-\alpha^2}b^{(n-1)\gamma}r_+^{(n-1)(1-\gamma)+1}.
\end{equation} 
We point out here that the thermodynamic volume derived for the linear fields case \cite{Stetsko_EMYMD20} is completely identical to the given above expression. Similarly as we did in our earlier works \cite{Stetsko_EYM20,Stetsko_EMYMD20,Stetsko_EnYM20} we assume that the nonabelian field parameter $\bar{q}$ might be varied and thus we introduce thermodynamic value corresponding to it. Namely we introduce the Yang-Mills charge which is defined as follows \cite{Stetsko_EnYM20}:
\begin{equation}\label{YM_charge}
\bar{Q}=\frac{1}{4\pi\sqrt{(n-1)(n-2)}}\int_{\Sigma}d^{n-1}\chi J(\Omega)\left(Tr(F^{(a)}_{\mu\nu}F^{(a)}_{\mu\nu})\right)^{\frac{p}{2}}=\frac{\omega_{n-1}}{4\pi}((n-1)(n-2))^{\frac{p-1}{2}}\bar{q}^p,
\end{equation} 
where $J(\Omega)$ is the Jacobian in the spherical coordinates and the integral is taken over a hypersphere enclosing the black hole. Since the introduced Yang-Mills charge is supposed to be a thermodynamic value then corresponding conjugate potential might be defined as follows:
\begin{equation}
\bar{U}=\left(\frac{\partial M}{\partial \bar{Q}}\right)_{S,Q,P}.
\end{equation}
Taking into account these new thermodynamic values we are able to write the so-called extended first law:
\begin{equation}\label{1st_law_e}
dM=TdS+VdP+\Phi_e dQ+\bar{U}d\bar{Q}.
\end{equation}
The extended phase space description also allows us to derive the Smarr relation, which might be written as follows:
\begin{equation}\label{smarr}
(n+\al^2-2)M=(n-1)TS+2(\al^2-1)VP+\frac{1}{s}(s(n-3)+1+\al\be)\Phi_{e}Q+(2p-1-\al\de)\bar{U}\bar{Q}.
\end{equation}
It is worth noting that the Smarr relation (\ref{smarr}) apart of fixed parameters such as the dimension $n$ or the coupling constants $\be$ and $\de$ also depends on the parameter $\al$ which is introduced in the relation (\ref{ansatz_R}) and might not be equal to any of the coupling constants. Actually, the parameter $\al$ defines the behaviour of the metric at large distances (at the infinity), and consequently it appears in the the Smarr relation. 
 
In the extended thermodynamics framework we are able to define the equation of state which can be represented in the following form:
\begin{gather}
\nonumber P=\frac{(n-1)}{4(1+\al^2)r_+}T-\frac{(n-1)}{16\pi}\left(\frac{n-2}{1-\al^2}b^{-2\g}r^{2(\g-1)}_{+}\left(1+\frac{2p(1-\al^2)\left((n-1)(n-2)\right)^{p-1}}{\al^2-2p+\al\de}\bar{q}^{2p}\times\right.\right.\\\left.\left.b^{-2(2p+\k_1)\g}r^{-4p(1-\g)+2\k_1\g}_{+}\right)-\frac{2^{s}s(2s-1)}{s(n-1)+\al\be-(2s-1)\al^2}q^{2s}b^{\frac{2(\k_2-s(n-1))\g}{2s-1}}r^{-2\frac{(s(n-1)(1-\g)+\k_2\g)}{2s-1}}_{+}\right).\label{EOS_1}
\end{gather}
Here we note that if we take $p=s=1$ and impose $\al=\de$ the equation of state is reduced to the corresponding equation for linear gauge fields derived in our recent work \cite{Stetsko_EMYMD20}. If we suppose that $q=0$ and $\al=\de$ then we arrive at the relation obtained in our study \cite{Stetsko_EnYM20} and when in addition we put $p=1$ the equation of state derived in \cite{Stetsko_EYM20} is recovered. The upper equation is supposed to be an analog of the Van der Waals equation of state, although there might be some sort of discontent since the obtained values for the pressure $P$ and the temperature $T$ are rather geometrical ones. Physical values are introduced as follows:
\begin{equation}
[P]=\frac{\hbar c}{l^{n-1}_{Pl}}P, \quad [T]=\frac{\hbar c}{k}T,
\end{equation}
where $l_{Pl}$ is the Planck length in $n+1$ dimensional space-time. Having used these physical values one might define also a specific volume, which would be proportional to the product $l^{n-1}_{Pl}r_+$, but since all these physical values are simply proportional to their geometrical counterparts the latter ones might be retained. For convenience we rewrite the equation of state (\ref{EOS_1}) in the following form:
 \begin{gather}
\nonumber P=\frac{T}{v}-\frac{(n-1)}{16\pi}\left(\frac{n-2}{1-\al^2}b^{-2\g}(\k v)^{2(\g-1)}\left(1+\frac{2p(1-\al^2)\left((n-1)(n-2)\right)^{p-1}}{\al^2-2p+\al\de}\bar{q}^{2p}\times\right.\right.\\\left.\left.b^{-2(2p+\k_1)\g}(\k v)^{-4p(1-\g)+2\k_1\g}\right)-\frac{2^{s}s(2s-1)}{s(n-1)+\al\be-(2s-1)\al^2}b^{\frac{2(\k_2-s(n-1))\g}{2s-1}}(\k v)^{-2\frac{(s(n-1)(1-\g)+\k_2\g)}{2s-1}}\right),\label{eos_2}
\end{gather}
where we have introduced new ``volume'' parameter $v=4(1+\alpha^2)r_+/(n-1)$ and $\k=(n-1)/(4(1+\alpha^2))$.
Due to full analogy between the written above equation of state (\ref{eos_2}) and van der Waals equation we investigate the equation of state (\ref{eos_2}) in the same manner as it is performed for the Van der Waals equation. In particular, the equation (\ref{eos_2}) might have an inflection point which is defined as follows:
\begin{equation}
\left(\frac{\partial P}{\partial v}\right)_T=0, \quad \left(\frac{\partial^2 P}{\partial v^2}\right)_T=0.
\end{equation}
The written above relation gives rise to the following relation for the the critical volume $v_c$:
\begin{gather}
\nonumber 1+\frac{2p(\al\de-2p)(2\al\de-4p+1+\al^2)}{(\al^2-2p+\al\de)}\left((n-1)(n-2)\right)^{p-1}\bar{q}^{2p}b^{-2(2p-1+\k_1)\g}(\k v_c)^{2(1-2p)(1-\g)+2\k_1\g}-\\\frac{2^{s}s(s(n-1)+\al\be)(2s(n-1)+2\al\be-(2s-1)(1+\al^2))}{(2s-1)(n-2)(s(n-1)+\al\be-(2s-1)\al^2)}q^{2s}b^{\frac{2(\k_2-s(n-1))\g}{2s-1}+2\g}(\k v_c)^{\frac{2(s(n-1)(\g-1)-\k_2\g)}{2s-1}+2(1-\g)}=0.\label{vc_eq}
\end{gather}
From the evident form of the equation (\ref{vc_eq}) it follows that it is not possible to find its general analytical solution, the critical volume might be derived only numerically. We point out here that this peculiarity is not a specific feature of the nonlinear gauge fields, similar situation takes place even for the linear gauge fields \cite{Stetsko_EMYMD20}. For particular cases when one of the gauge fields is not taken into consideration the relation (\ref{vc_eq}) can be solved analytically. Nevertheless, the expressions for the critical temperature $T_c$ and pressure $P_c$ might be written in general case. Thus, we obtain:
\begin{gather}
\nonumber T_c=\frac{(n-1)(n-2)b^{-2\g}\k^{2(\g-1)} v_c^{2\g-1}}{4\pi(1+\al^2)(2s(n-1)+2\al\be-(2s-1)(1+\al^2))}\left(\frac{s(n-3)+\al\be+1}{(1-\al^2)}-2p((n-1)(n-2))^{p-1}\times\right.\\\left.\frac{(\al\de-2p)(s(n-1)+\al\be+(2s-1)(\al\de-2p))}{\al^2-2p+\al\de}\bar{q}^{2p}b^{-2(2p-1+\k_1)\g}(\k v_c)^{2(1-2p)(1-\g)+2\k_1\g}\right);
\end{gather}
\begin{gather}
\nonumber P_c=\frac{(n-1)(n-2)b^{-2\g}(\k v_c)^{2(\g-1)}}{16\pi(1+\al^2)(s(n-1)+\al\be)}\left(s(n-1)+\al\be-2s+1-2p((n-1)(n-2))^{p-1}\times\right.\\\left.\frac{(2\al\de-4p+1+\al^2)(s(n-1)+\al\be+(2s-1)(\al\de-2p))}{\al^2-2p+\al\de}\bar{q}^{2p}b^{-2(2p-1+\k_1)\g}(\k v_c)^{2(1-2p)(1-\g)+2\k_1\g}\right).
\end{gather}
For the linear field case $s=p=1$ and if $\al=\de$ the latter two relations are reduced to the corresponding critical values derived in our earlier work \cite{Stetsko_EMYMD20}. Similarly for other particular cases the corresponding critical values are recovered  \cite{Stetsko_EYM20,Stetsko_EnYM20}.

\begin{figure}
\centerline{\includegraphics[scale=0.33,clip]{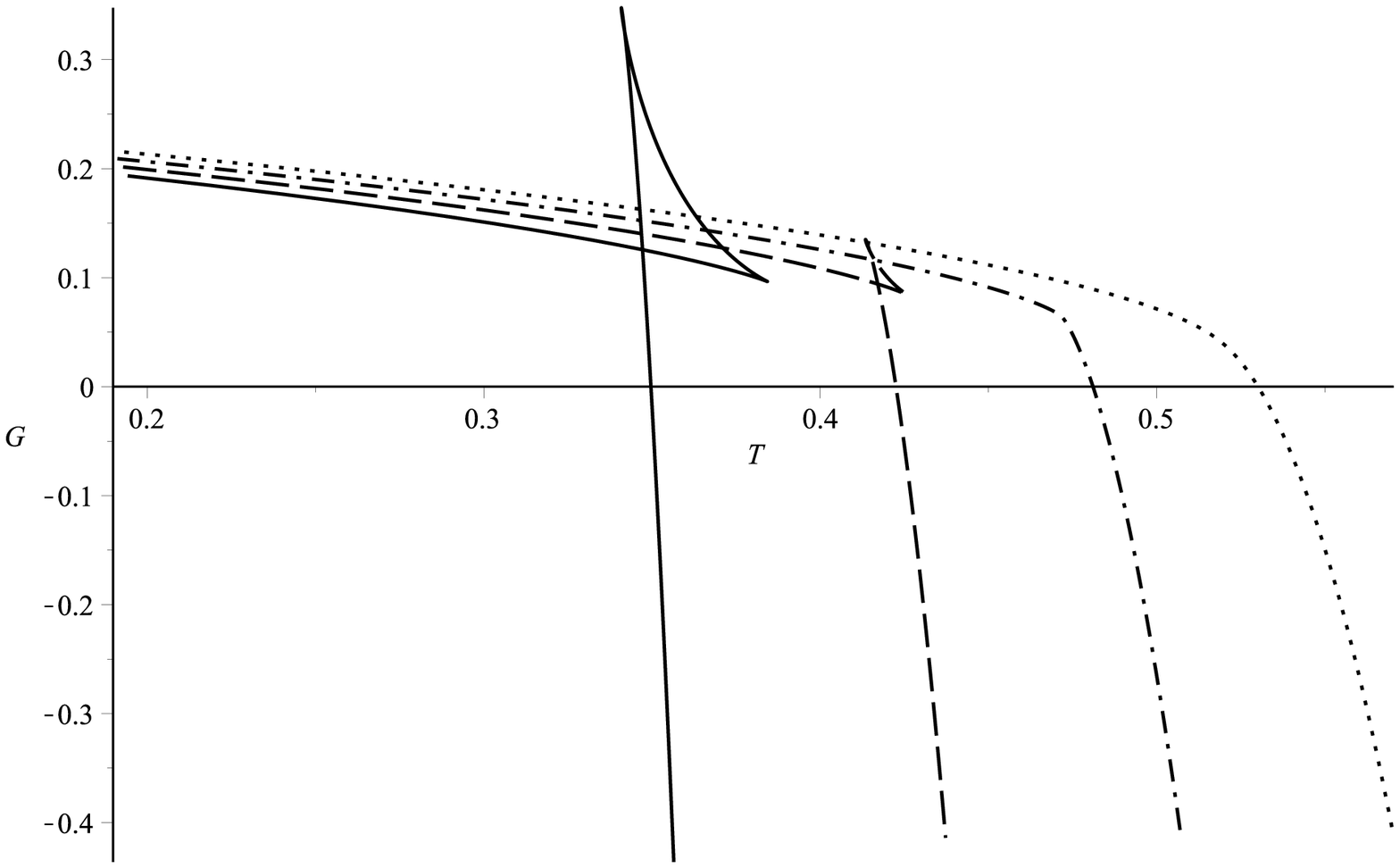}\includegraphics[scale=0.33,clip]{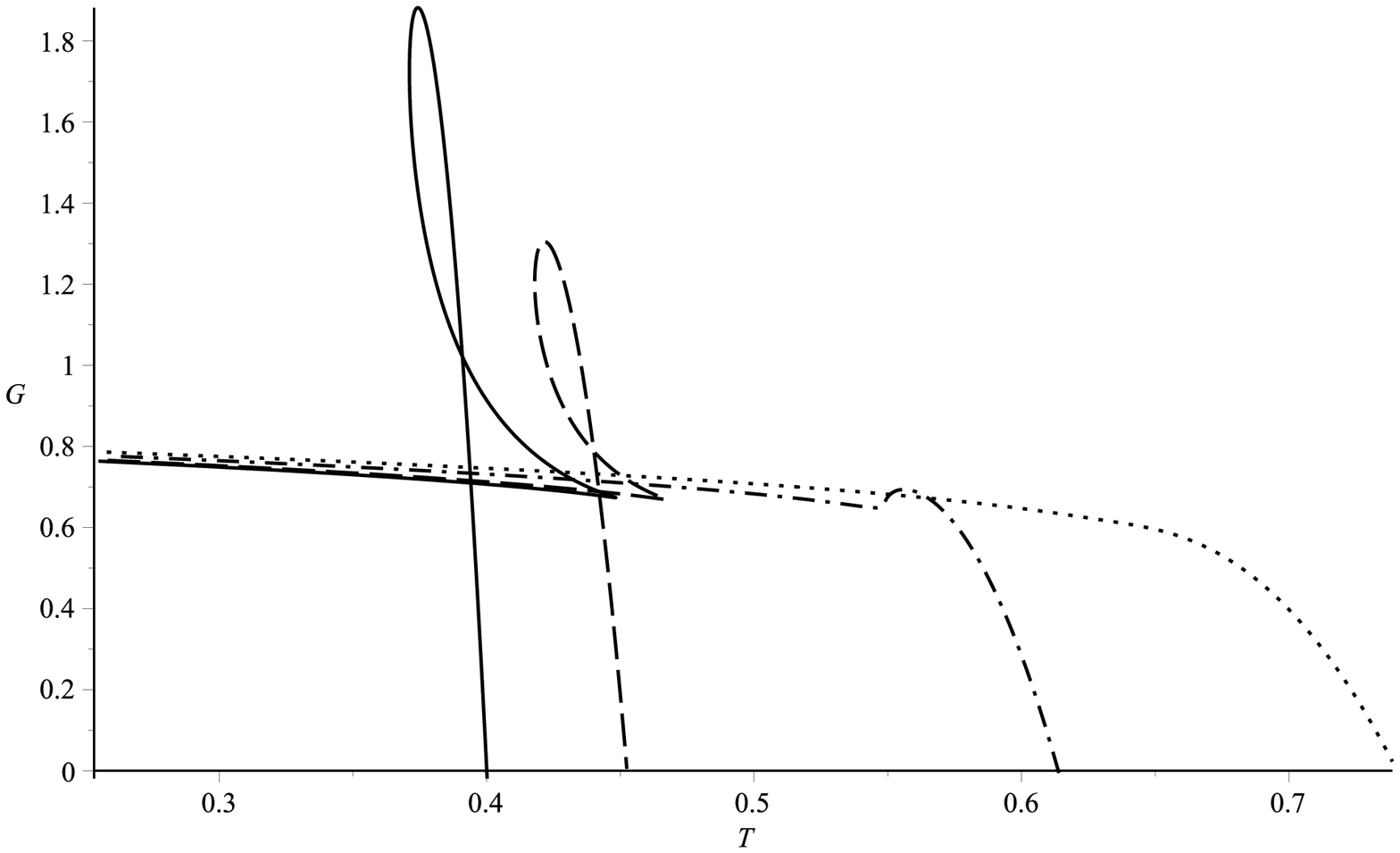}}
\caption{Gibbs free energy for some values of pressure if the system shows standard swallowtail behaviour ($\al=0.1$ is relatively small) and  additional loop becomes visible ($\al=0.4$ for some intermediate value of this parameter). For both graphs we have taken $n=5$, $\be=0.12$, $\de=0.1$, $\bar{q}=q=0.2$, for the left graph we have chosen $p=s=1$ and for the right one $p=1.5$ and $s=1.3$. For both graphs the dash-dotted lines correspond to the critical pressures, the dotted lines correspond to the pressures above the critical pressure and solid and dashed curves correspond the pressures below the critical one, their values are chosen just to reflect the typical behaviour of the function $G=G(T)$.}\label{Gibbs_gr_1}
\end{figure}

\begin{figure}
\centerline{\includegraphics[scale=0.33,clip]{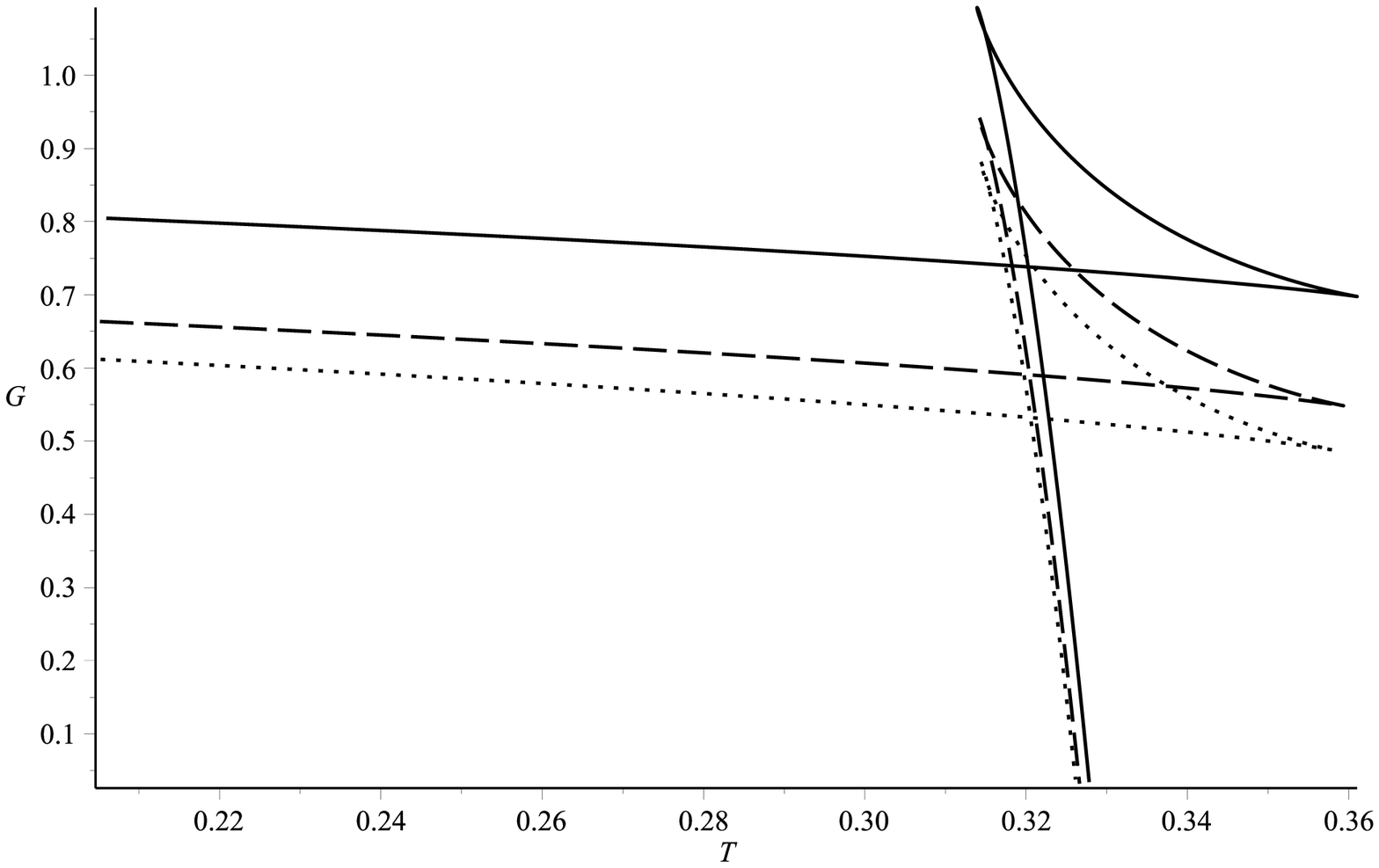}\includegraphics[scale=0.33,clip]{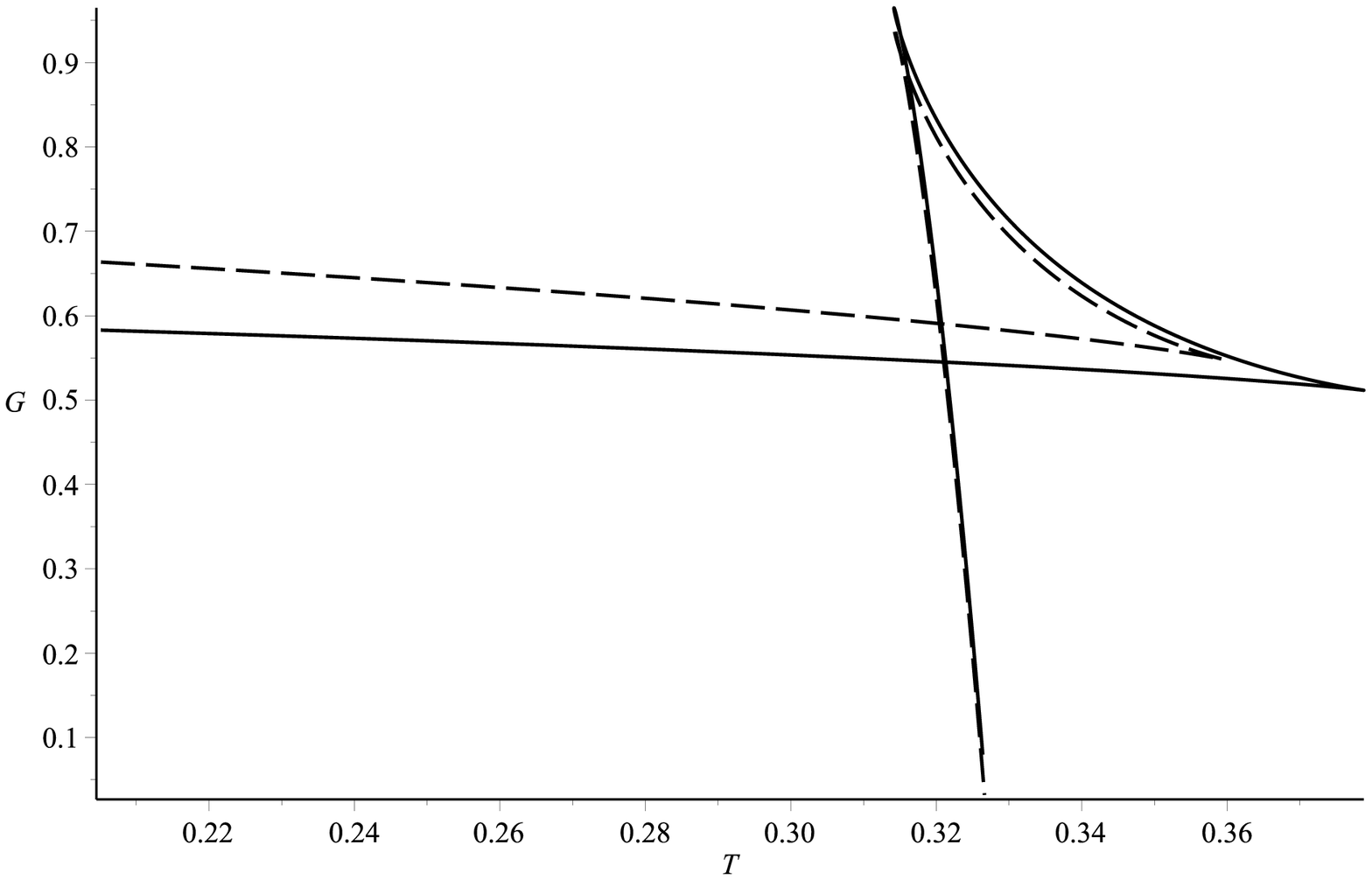}}
\caption{Influence of the variation of nonlinearity parameters $p$ (the left graph) and $s$ (the right one) on the Gibbs free energy. For both graph we have chosen $P=0.1$, $n=5$, $\al=0.1$, $b=1$, $\be=0.12$, $\de=0.1$, $\bar{q}=q=0.2$. For the left graph the solid, dashed and dotted curves correspond to $p=1.4$, $p=1.5$ and $p=1.6$ respectively and for all of them $s=1$. For the right graph dashed and solid curves correspond to $s=1$ and $s=2$ respectively and $p=1.5$. }\label{Gibbs_gr_2}
\end{figure}

Gibbs free energy is known to be very useful for studies of critical behaviour of a system. It was investigated in various papers where the extended thermodynamics for dilaton black holes was examined \cite{Hendi_PRD15,Dehyadegari_PRD17, Stetsko_EYM20,Stetsko_EMYMD20,Stetsko_EPJC19}. For our system we obtain:
\begin{gather}\label{gibbs_en}
\nonumber G=\frac{(1+\al^2)\omega_{n-1}}{16\pi}b^{(n-1)\g}r^{(n-1)(1-\g)}_{+}\left(\frac{n-2}{n+\al^2-2}b^{-2\g}r^{2\g-1}_{+}+\frac{2p(n-1)^{p-1}(n-2)^{p}(4p-2\al\de-\al^2-1)}{(\al^2-2p+\al\de)(n+\al^2-4p+2\al\de)}\times\right.\\\left.\nonumber\bar{q}^{2p}b^{-2(2p+\k_1)\g}r^{1+2\k_1\g-4p(1-\g)}_{+}+\frac{16\pi(\al^2-1)}{(n-1)(n-\al^2)}Pr_{+}+\right.\\\left.\frac{2^{s}s(2s-1)((2s-1)(n-\al^2-1)+n+2\al\be-2s)q^{2s}}{(s(n-1)+\al\be-(2s-1)\al^2)(n+2\al\be-2s(1+\al^2)+\al^2)}b^{\frac{2(\k_2-s(n-1))\g}{2s-1}}r^{1+\frac{2(s(n-1)(\g-1)-\k_2\g)}{2s-1}}_{+}\right).
\end{gather}
It should be pointed out here that written above relation might be considered as a generalization of the result, obtained for linear gauge fields \cite{Stetsko_EMYMD20}, or in the absence of the Maxwell field for nonlinear Yang-Mills case \cite{Stetsko_EnYM20}. We also note that from the qualitative point of view the behaviour of the Gibbs free energy (\ref{gibbs_en}) is very similar to the corresponding behaviour of the Gibbs free energy derived for linear fields \cite{Stetsko_EMYMD20}, its behaviour is shown on the Figure [\ref{Gibbs_gr_1}]. The Gibbs free energy as a function $G=G(T,P)$ would be a monotonous function if one of its arguments either temperature $T$ or the pressure $P$ is above its critical value and what is clearly shown on the Figure [\ref{Gibbs_gr_1}]. Below the critical values the behaviour of the Gibbs free energy becomes nonmonotonous with specific swallowtail behaviour when the dilaton parameter $\al$ is relatively small. If this parameter goes up the swallowtail behaviour is modified by a specific loop. We also point out here that in our previous works we assumed that the parameter $\al$ in the radial function (\ref{ansatz_R}) was equal at least to one of the coupling constants between dilaton and gauge fields \cite{Stetsko_EYM20,Stetsko_EMYMD20,Stetsko_EnYM20}, and consequently we made a conclusion that if the coupling parameter increases then the specific loops become visible. Now we conclude that  if $\al\neq\be$ and $\al\neq\de$ the appearance of the closed loops depends on the value of the parameter $\al$. It is known that swallowtail behaviour of the Gibbs free energy implies that there is a phase transition of the first order \cite{Kubiznak_JHEP12} what was shown to take place for various black holes \cite{Kubiznak_CQG17}. Additional loop on the swallowtail emerges for dilaton black holes \cite{Dehyadegari_PRD17,Stetsko_EYM20,Stetsko_EPJC19}, it was shown to be related to the phase transition of the zeroth order which might appear for dilatonic black holes. The phase transition of the zeroth order occurs immediately below the critical point, but when the loop becomes closed as a consequence the Gibbs free energy turns to be continuous, therefore below this point the phase transition will be of the first order, as it usually takes place for various types of black holes. We also note that variation of nonlinearity parameters $p$ and $s$ does not change qualitative features of the Gibbs free energy what is reflected on the Figure [\ref{Gibbs_gr_2}]. 

We would also like to draw some attention to the critical point. Usually it is the point of the second order phase transition, since the second derivatives of the Gibbs free energy $G=G(T,P)$ are divergent here. But, as it was noted in our earlier work \cite{Stetsko_EnYM20} there are some peculiarities here in comparison with the standard phase transition of the second order, namely the so-called Prigogine-Defay ratio is not equal to one, as it has to be for the standard phase transition of the second order. Similarly we introduce the Prigogine-Defay ratio here, which is defined as follows:
\begin{equation}\label{PD_rat}
\Pi=\frac{\Delta C_P\Delta\kappa_T}{VT(\Delta\tilde{\al})^2},
\end{equation} 
where $C_P=T(\partial S/\partial T)_P$ is the heat capacity, $\k_T=-1/V(\partial V/\partial P)_T$ is the isothermal compressibility and $\tilde{\al}=1/V(\partial V/\partial T)_P$ is the volume extension coefficient and $\Delta$ denotes the jump of the corresponding values at the critical point. It might be shown easily that $\k_T=\tilde{\al}(\partial T/\partial P)_V$. Since the heat capacity $C_P$ and the volume extension coefficient $\tilde{\al}$ have the same character of divergence at the critical point it means that at the critical point the Prigogine-Defay ratio (\ref{PD_rat}) is finite. After calculation of derivatives we obtain:
\begin{equation}\label{PD_dil}
\Pi=\frac{n-\al^2}{n+\al^2}.
\end{equation}
From the latter relation it follows that if $\al\to 0$ then $\Pi\to 1$, but for any nonzero value of $\al$ the ratio $\Pi<1$, and it decreases if the parameter $\al$ goes up, this fact might be related to the appearance of zeroth order phase transition and consequently to specific loops at the $G=G(T)$ or $G=G(P)$ diagrams. In our case the Prigogine-Defay ratio is less then one and we might suppose that at the critical point there is a glass-type phase transition, although for the latter one usually the ratio $\Pi>1$ \cite{Schmelzer_JCP06}. We also point out that the Prigogine-Defay ratio (\ref{PD_dil}) has some ``universal'' feature, namely it does not depend on the integration constants which are present in the metric functions and which are inherited by the corresponding thermodynamic values, it depends on the dimension $n$ and the parameter $\al$ which might be related to the dilaton coupling constants. We also note that the final form for the Prigogine-Defay ratio (\ref{PD_dil}) coincides with the corresponding expression obtained for a simpler model, namely the black hole in Einstein-dilaton-power-Yang-Mills theory. 
\section{Conclusions}
In this work we have considered Einstein-dilaton theory with nonlinear abelian and nonabelian gauge fields with nonlinearities of power-law type. We have obtained static spherically symmetric black hole solution in this framework. The obtained solution might be considered as a direct generalization of the solution derived in our previous work \cite{Stetsko_EMYMD20} for Einstein-Maxwell-Yang-Mills-dilaton theory  with standard linear gauge fields. It should be pointed out that some important features of the linear gauge fields solution \cite{Stetsko_EMYMD20} take place here for nonlinear fields case, namely the black hole might have two horizons, what is clearly demonstrated by the Figure [\ref{W_graph}] and the outer one is the event horizon. Increasing of the corresponding gauge charge might give rise to merging of the horizons and in this case the so-called extreme black hole appears. With the following increase of the gauge charge the extreme black hole gets transformed into a naked singularity. In contrast to the linear fields case where the leading term for small distances was related to the Maxwell field, here the leading term at small distances is defined by the value of nonlinearity parameters $p$ or $s$ and dimension of space $n$ we examine. In the limit of linear gauge fields ($p=s=1$), the leading term for small distances as it is known is caused by the Maxwell field term if $n>3$, and if $n=3$ both gauge fields might have very close orders depending on the value of the parameter $\al$. At large distances (asymptotic infinity) the metric is not asymptotically flat, but it is not exactly of de Sitter or anti-de Sitter type, the character of dependence is completely determined by the parameter $\al$ and if the latter goes to zero, the metric tends to the dS- or AdS-types. We conclude that the parameter $\al$ plays very important role in our solution, namely it defines the behaviour of the metric function $W(r)$ at the infinity and actually it might be treated as a dilaton field coupling parameter. 

We have also examined thermodynamics of the obtained black hole. At first we derived and examined the behaviour of the black hole's temperature $T$ as a function of the horizon radius $r_+$. From the qualitative point of view it is similar to the corresponding behaviour of the temperature for linear field case, obtained in our previous work \cite{Stetsko_EMYMD20} or other particular cases. Namely, the temperature is nonmonotonous for some intermediate values of the horizon's radius and this fact implies some sort of criticality. The influence of the variation of various parameters on the temperature is shown on the Figures [\ref{temp_graph}]-[\ref{temp_gr_2}], from the given graphs it is easy to conclude that variation of the nonlinearity parameters $p$ and $s$ has its impact for relatively small radii of the horizon $r_+$, variation of the cosmological constant affects on the behaviour of the temperature for large radii of the horizon and finally the variation of the parameter $\al$ has its influence on all the values of the radius of the horizon. We have also derived the first law of black hole's thermodynamics. One of the most important issues in black holes' thermodynamics is the investigation of thermal stability, it might be inferred from the examination of the heat capacity. Here we point out that qualitative features of the heat capacity are the same  as for linear gauge fields case at least when the nonlinearity parameters $p$ and $s$ do not differ from one substantially \cite{Stetsko_EMYMD20}.

The other interesting issue in black holes' thermodynamics is the so-called extended phase space approach. It is based on the interpretation of the cosmological constant $\L$ as a thermodynamic variable related to the pressure \cite{Kastor_CQG09,Kubiznak_CQG17}. The extended phase space approach allowed us to derive the equation of state for the black hole (\ref{EOS_1}) which is usually considered as an analog of the van der Waals equation of state for ordinary liquid-gas system. Investigation of the equation of state allows to derive critical values for the temperature $T_c$, the pressure $P_c$ and the volume $v_c$, although the latter is not obtained in evident form, because of the complicated structure of the equation (\ref{vc_eq}). The extended thermodynamics concept also allowed us to derive the Smarr relation (\ref{smarr}). 

Within extended thermodynamics concept we have calculated the Gibbs free energy (\ref{gibbs_en}). Since the complicated structure of the Gibbs free energy we have shown its behaviour graphically on the Figures [\ref{Gibbs_gr_1}]-[\ref{Gibbs_gr_2}]. The Gibbs free energy shows specific swallow-tail behaviour for relatively small values of the parameter $\al$ below the critical pressure (or temperature). When they are above the critical values the Gibbs free energy turns to be monotonous. The swallow-tail behaviour of the Gibbs free energy is a characteristic feature of the first order phase transition. If the parameter $\al$ increases the typical swallow-tail form becomes modified by a specific loop. This modification implies that here in addition to the first order phase transition the zeroth order phase transition appears. The latter takes place in the domain where the loop is not closed yet. Here we also note that the zeroth order phase transition takes place for black holes with dilaton fields \cite{Dehyadegari_PRD17,Stetsko_EYM20,Stetsko_EMYMD20,Stetsko_EPJC19}. It is worth pointing out that variation of the nonlinearity parameters does not affect on qualitative character of the  Gibbs free energy at least in the
chosen domain of variation. Finally, to comprehend the thermal behaviour at the critical point better we have calculated the Prigogine-Defay ratio (\ref{PD_rat}) and the resulting expression (\ref{PD_dil}) differs from one as it usually takes place for the phase transition of the second order. So the phase transition at the critical point is not exactly of the second order, it is closer to the phase transition of glass type, even though for the phase transition of the glass type it is normally  $\Pi>1$, whereas in our case $\Pi<1$.

\section{Acknowledgments}
This work was partly supported by Project FF-83F (No. 0119U002203) from the Ministry of Education and Science of Ukraine.  

\end{document}